\documentclass[referee]{RAA}
\usepackage{graphicx, times}
\usepackage{multirow}

\begin{document}
\title{A Statistical Definition of Image Resolution Based on the Correlation of Pixels}
   \volnopage{Vol.0 (200x) No.0, 000--000}      
   \setcounter{page}{1}          
\author{Jian-Feng Zhou
	\inst{1,2}
	}
\institute{Department of Engineering Physics and Center for Astrophysics,
Tsinghua University, Beijing 100084, China; {\it zhoujf@tsinghua.edu.cn}\\
\and
Key Laboratory of Particle \& Radiation Imaging (Tsinghua University),
Ministry of Education, Beijing 100084, China.
}

\date{Received~~2015 month day; accepted~~2015~~month day}

\abstract{
Resolution, usually defined by the Rayleigh criterion or the Full Width at Half Maximum of 
a Point Spread Function, is a basic property of an image. Here, we present a new statistical definition of 
image resolution based on the cross-correlation properties of the pixels in an image. 
It is shown that the new definition of image resolution depends not only on the PSF of an imaging device,
but also on the signal-to-noise ratio of the data and on the structures of an object. In an
 image, the resolution does not have to be uniform. Our new definition
is also suitable for the interpretation of the result of a deconvolution.
We illustrate this, in this paper, with a Wiener deconvolution. It is found that weak 
structures can be extracted from low signal-to-noise ratio data, but with low resolution; a high-resolution image was 
obtained from high signal-to-noise ratio data after a Wiener deconvolution. 
The new definition can also be used to compare various deconvolution algorithms on their processing effects, such
as resolution, sensitivity and sidelobe level, etc.
}

\authorrunning{J.-F. Zhou}
\titlerunning{A Statistical Definition of Image Resolution}

\maketitle

\section{Introduction}
Resolution, or spatial resolution, or angular resolution, describes the ability of an imaging 
device to distinguish small details of an object. 
In optical devices such as an optical or radio telescope, a microscope etc.,  Airy pattern 
is the description of the best focused spot of light that a perfect lens with a circular 
aperture can make, limited by the diffraction of light. 
The resolution of such imaging systems can be estimated by the Rayleigh criterion: 
two point sources are regarded as just resolved when
the principal diffraction maximum of one image coincides with the first minimum of the other (\cite{born1999principles}).
For a general imaging system, like a coded-mask, an interferometry, etc., 
the FWHM (Full Width at Half Maximum) of its PSF (Point Spread Function) 
is also widely used for defining the resolving power of the system.

In practice, one is  interested more in the resolution of an image itself than in that of an imaging system. 
An observed image is related to the properties of its imaging system, but also to the intensity 
distribution of an object, and to additive noise. 
Therefore, the resolution of an image is not only influenced by the PSF of an imaging system, 
but also by the structures of the object and noise. It is useful to find a new definition that can be 
used for directly estimating the resolution of an observed image.

Super-resolution images can be obtained through various methods.  Irani and Peleg (\cite{irani1991improving}) proposed an 
algorithm based on image registration where a series of low-resolution images were used and their relative 
displacements were known accurately.  Even for one single observed image blurred by an imaging system, 
deconvolution or blind deconvolution can be used for improving the resolution (\cite{conchello1998superresolution}). Fourier ring correlation (FRC) (\cite{saxton1982correlation}; \cite{nieuwenhuizen2013measuring}; \cite{banterle2013fourier}), which can be computed directly from an image, was applied to estimate the resolution of optical nanoscopy (or super-resolution microscopy).   

In principle, a proper quantitive definition of resolution needs detailed information about the relationship of the pixels of an image. So far, there is still no such kind of definition.
In this paper, a new definition of  image resolution based on the statistical properties of observed images is 
proposed.
In section \ref{sec2}, some basic factors that influence image resolution are introduced. The 
relationship between the correlation of pixels and image resolution is indicated in section \ref{sec3}, and then
a new definition of image resolution is presented there. 
In section \ref{sec4}, Wiener deconvolution is used to demonstrate  super-resolution and dynamic resolution
 under different  signal-to-noise ratio conditions . Finally, a discussion and conclusions are
 presented in section \ref{sec5} and \ref{sec6}.

\section{Factors Related to Image Resolution} \label{sec2}
A linear imaging system can be described by the following equation 
\begin{equation} \label{imageconv}
\mathbf{d} = \mathbf{o} \otimes \mathbf{p} + \mathbf{n},
\end{equation}
where $\mathbf{d}$ is an observed image, $\mathbf{o}$ is  an object or a scene, 
$\mathbf{p}$ is the PSF of the system, $\mathbf{n}$ is additive noise, and $\otimes$ denotes convolution.

In the classical definition of image resolution, only the properties of a PSF is considered,
whereas a real observed image is not only influenced by the PSF,
but also by the structures of an object and noise.

\subsection{Point Spread Function}
As mentioned in the introduction, the Rayleigh criterion or the FWHM of a PSF can be used 
as a definition of image resolution. But these definitions are not suitable for all kinds of PSFs.  
Some PSFs' shapes  are rather strange and complex. They often have negative minimum, which 
has no physical meaning, and strong sidelobes that make central beam insignificant.
For example, the PSF of a coded-mask imaging system usually has
big  sparse sidelobes (\cite{ubertini2003ibis}).  
Also, in speckle interferometry, although high resolution information is preserved,  
the PSF in each frame with short exposure has large spread area  and a strong fluctuating structure due to atmospheric turbulence, with no obvious main beam (\cite{fried1966optical}, \cite{labeyrie1970attainment}). 

\subsection{Object Structures}
It is shown in (\ref{imageconv}) that the intensity distribution of object $\mathbf{O}$ 
has  an influence on the observed image. 

Due to the stochastic properties and to the particle nature of light, there is always fluctuation in an 
observed image: this is called shot noise (\cite{blanter2000shot}). The brighter the source, the stronger the noise. 
Such fluctuation can usually be described by Poisson distribution. If the photon number in a pixel
 is $N$, then the relative fluctuation level is proportional to $1/\sqrt{N}$.

\begin{figure}[!h]
\centering
\includegraphics[width=\textwidth, angle=0]{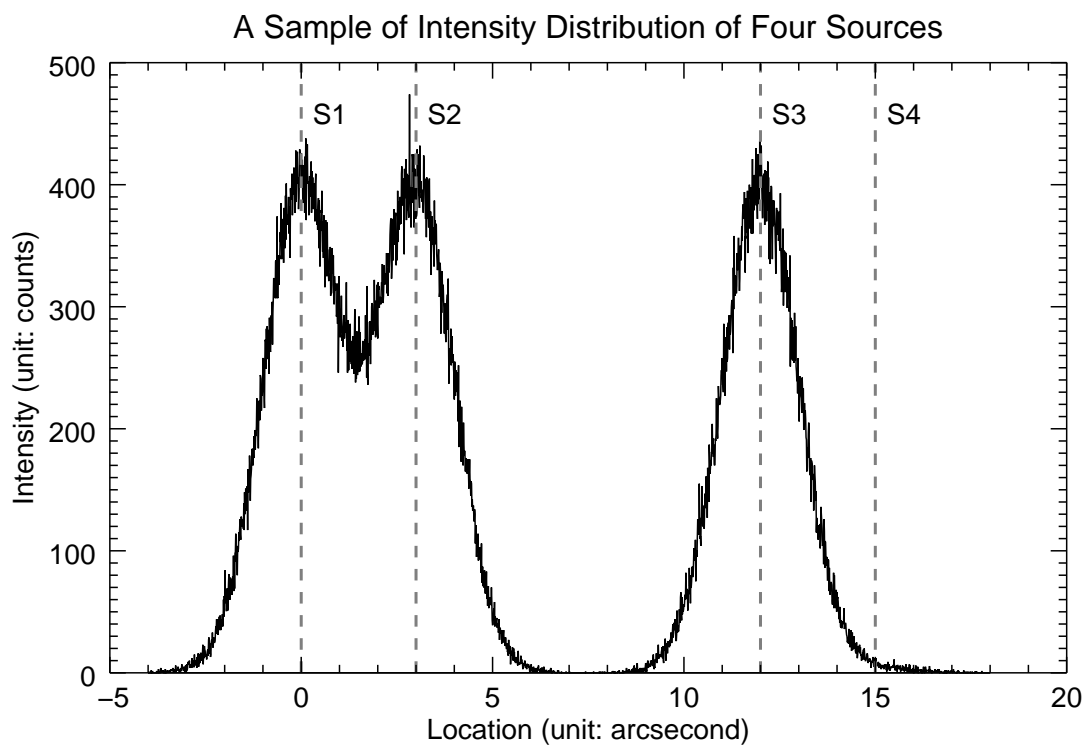}
\caption{Example showing how the structure of an object affects the resolution and detectability
when shot noise is taken into account. }
\label{fig01}
\end{figure}

In a real observed image, the exposure time is limited. 
Therefore, faint structures will generally be drown in the large fluctuations of nearby bright structures 
because of their large shot noise.

Let's use an example to show how the structure of an object  affects the resolution of an image (see Fig.\  \ref{fig01}).   
Here, a normal
Gaussian PSF with FWHM of 2.3 arcsecond is used. Four point sources are located at 
position 0, 3, 12 and 15 arcseconds and denoted by  S1, S2, S3 and S4 respectively.  
S1, S2 and S3 have an identical number of 
received photons, which is $10^5$. S4 is much weaker and has $10^3$ photons.
Since the distances between S1 and S2, as well as S3 and S4 are greater than the 
FWHM of the PSF, they should be resolvable under the classical definition of image resolution. 
Fig.\  \ref{fig01} clearly shows that S1 and S2 are resolved. However, S4 is shadowed by 
 its neighboring strong source S3.   

\subsection{Additive Noise}
An image's additive noise,  sometimes known as "dark shot noise" (\cite{macdonald2006digital}) or 
"dark-current shot noise" (\cite{janesick2001scientific}), is usually produced by the sensor and circuitry 
of a scanner or digital camera. 
This noise is an undesirable by-product of  the physical image capture process. 
In a dim scene, the additive noise (\cite{tuzlukov2010signal}) will be dominant and has an influence similar
to  that of shot noise .  It may conceal faint sources and  add spurious and extraneous information. 

\section{Correlation and Image Resolution} \label{sec3}
Correlation is a mathematical term that describes a relationship between two 
random variables.  There are several definitions of correlation. Among them is Pearson's definition (\cite{pearson1895note}), 
which captures the linear relationship between two random variables and is the most popular.

For an observed image, the counts of pixels are random variables whose statistics can be used to
study the resolution property of the image.   
The means of these variables match a distribution which can be described by (\ref{imageconv}).
The classical definitions of image resolution are relevant to this distribution. 
In this paper, quadratic statistics, and especially correlation coefficients are used for analyzing the resolution property
of an image.  

\subsection{Correlation Induced by Imaging Devices }
Signals emitted from different objects or from different parts of  the same object are usually
incoherent. This property can be mathematically described by Pearson's cross-correlation
coefficient. If $S_1$ and $S_2$ are two signals from different places, then the 
correlation coefficient (abbreviated as CC hereafter) between them is  $\rho(S_1,S_2)=0$.

For an imaging device, if the size of a pixel in its detector is much larger than
the FWHM of the PSF, then different pixels in a received image represent different
parts of the object or scene. In this situation, $\rho(R_1, R_2)=0$, where $R_1$ and $R_2$
represent the signals from two pixels, i.e., the signals from different pixels are
incoherent. If the size of a pixel is less than the FWHM of the PSF,
however, the signals from different pixels are generally coherent.

\begin{figure}[!h]
\centering
\includegraphics[width=\textwidth, angle=0]{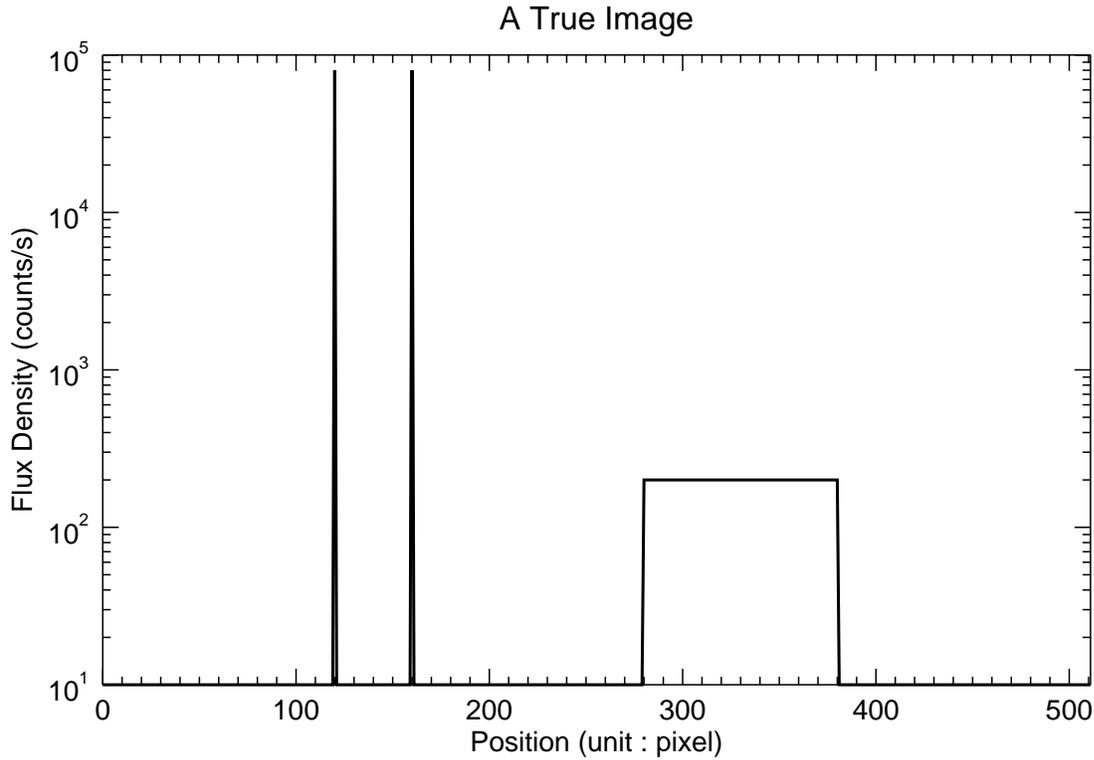}
\caption{Brightness distribution of a simulated object or true image. }
\label{fig02}
\end{figure}

Let's use a simulated one-dimension example to demonstrate the influence of an imaging device on the CC. 
The object (see Fig.\ \ref{fig02})  consists of two point sources and one 
extended source.
The point sources are located at pixels 120 and 140 respectively and have the same brightness of
80000 counts/s. The extended source resides in [280, 380] has a brightness of 200 counts/s.

\begin{figure}[!h]
\centering
\includegraphics[width=\textwidth, angle=0]{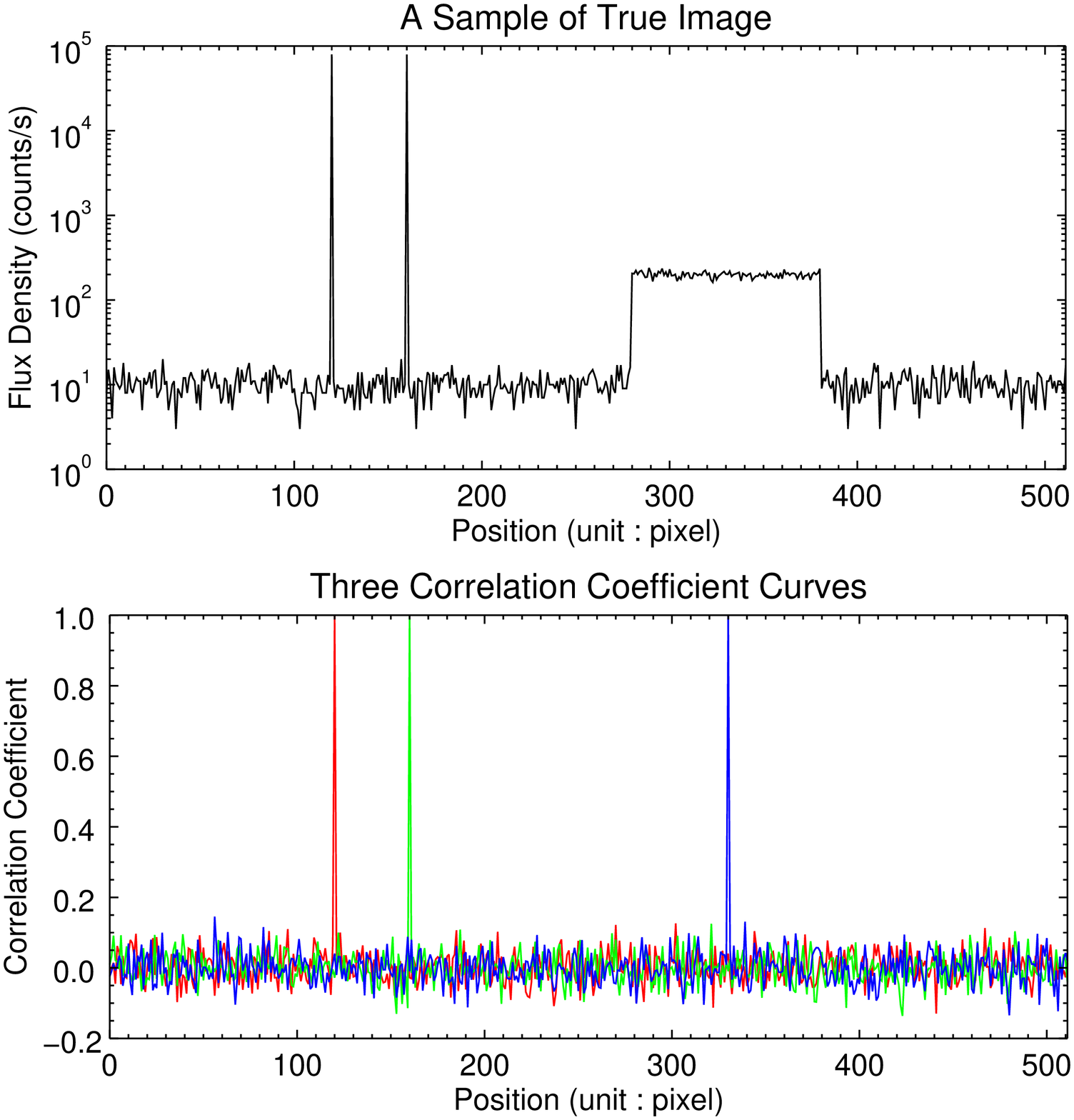}
\caption{One of the observed images, in a situation where the 
PSF is much smaller than
the pixels of detector (top plot). Three correlation coefficient curves  related
to pixels 120 (red), 160 (green) and 330 (blue) respectively  are shown in the bottom plot.}
\label{fig03}
\end{figure}

\begin{figure}[!h]
\centering
\includegraphics[width=8cm]{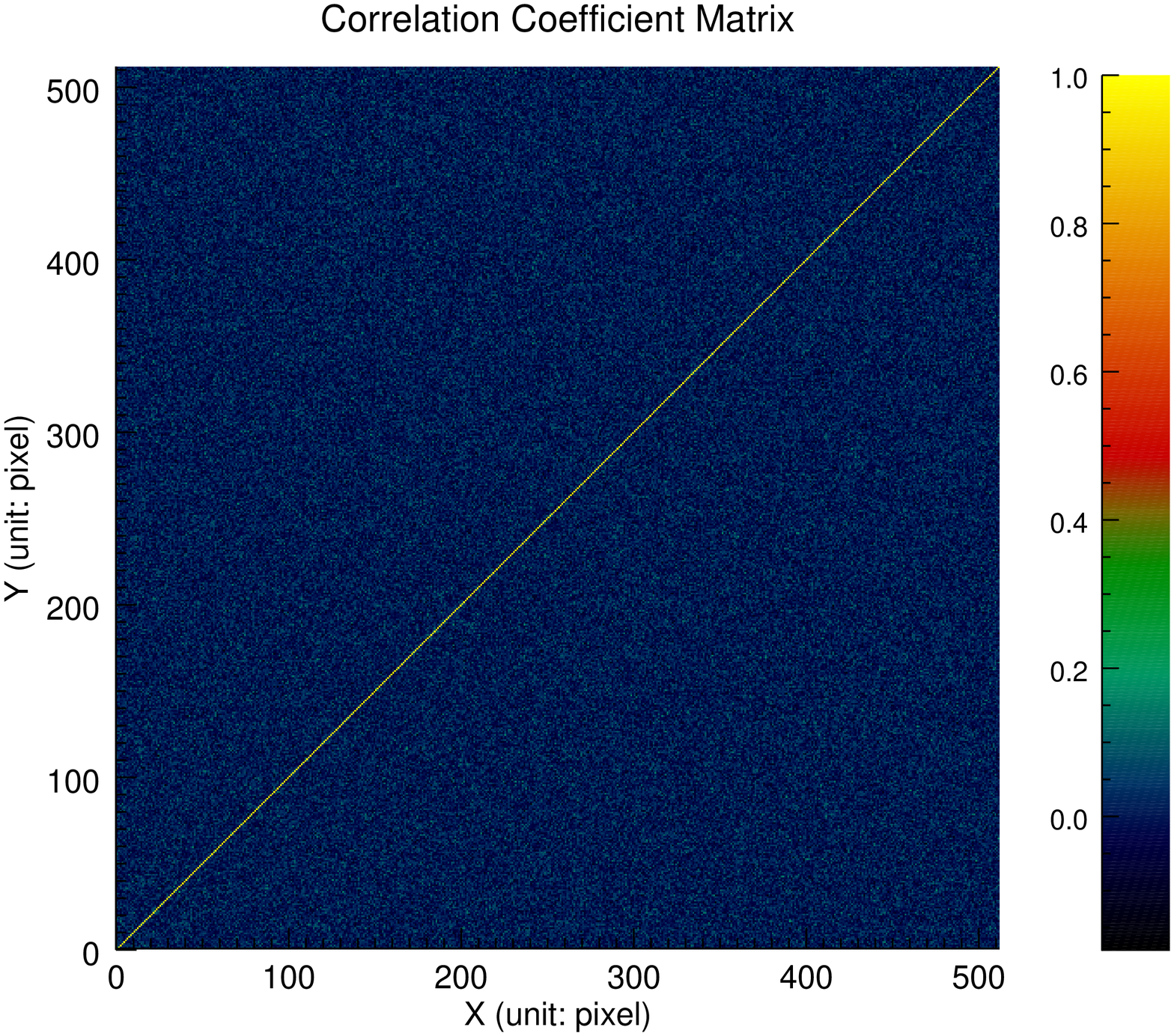}
\caption{The CC matrix of a series of observed images under a situation where the 
PSF is much smaller than
the pixels of the detector. }
\label{fig05}
\end{figure}

First, we choose a situation where the PSF is smaller than a pixel of the detector,
A series of simulated observed images are obtained. One sample where $1\,\mathrm{pixel}=1\,\mathrm{arcsecond}$ 
is shown
in  Fig.\ \ref{fig03} (top plot). The cross-correlation coefficient matrix
can be calculated and is shown in Fig.\ \ref{fig05}. Three coefficient curves
related to specific pixels are extracted from that matrix and shown in Fig.\ \ref{fig03} (bottom plot).

Fig.\ \ref{fig03} shows that all of three CC curves are essentially  $\delta$-functions, which means that
there is no correlation between any pair of pixels in the image.  The same is also true for the CC
matrix (see Fig.\ \ref{fig05}) where only diagonal elements are equal to 1 and other elements are close to 0.

\begin{figure}[!h]
\centering
\includegraphics[width=\textwidth]{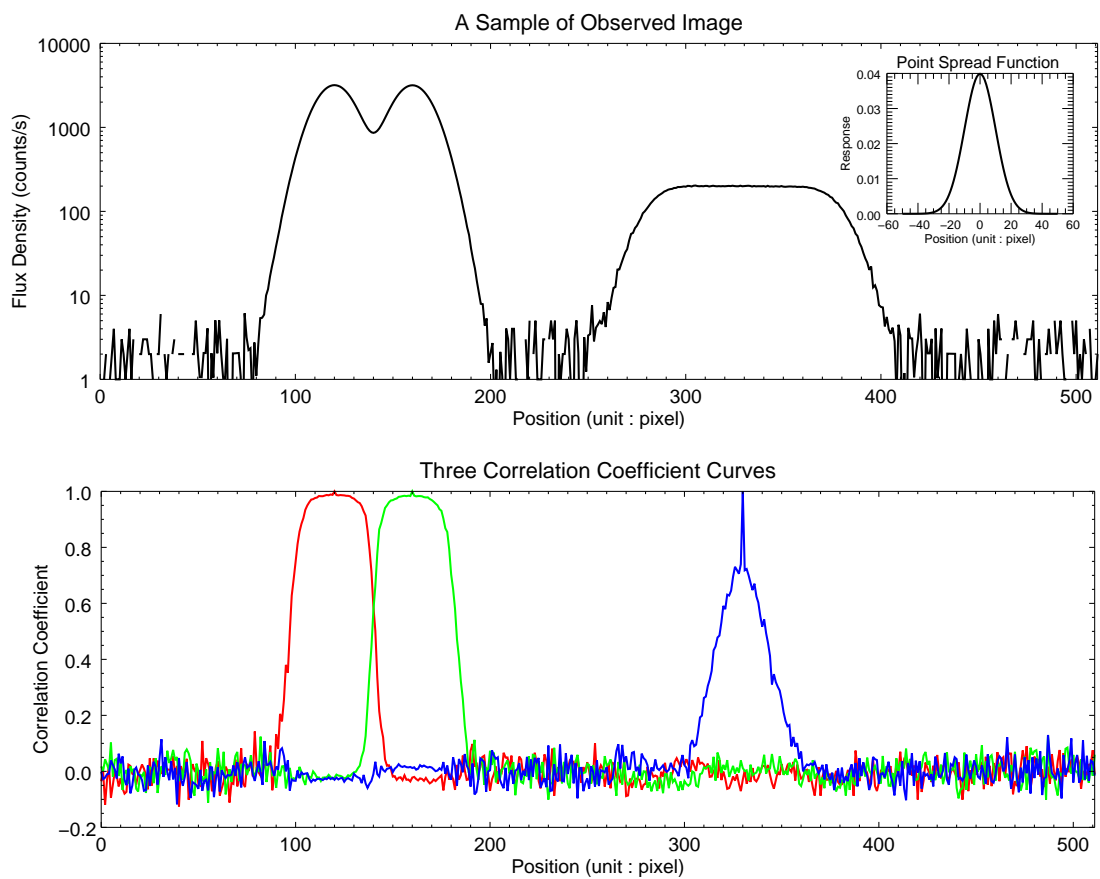}
\caption{One observed image, in a situation where the 
PSF is a Gaussian function, with $\sigma=10$ pixels (top plot). Three CC 
curves, for pixels 120 (red), 160 (green) and 330 (blue) respectively are shown in the bottom plot.}
\label{fig07}
\end{figure}

\begin{figure}[!h]
\centering
\includegraphics[width=\textwidth]{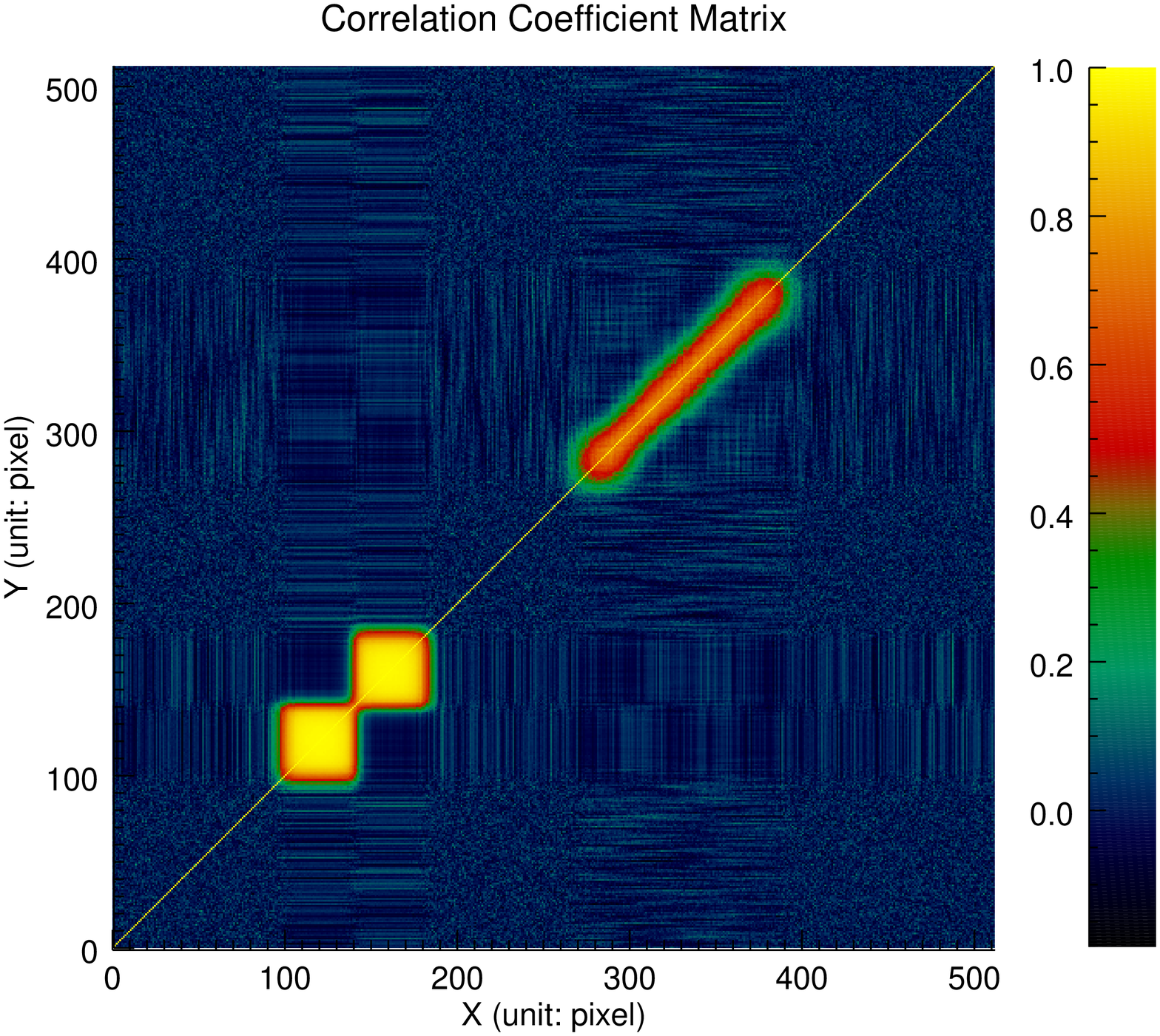}
\caption{The CC matrix of a series of observed images in a situation where the 
PSF is a Gaussian function, with $\sigma=10$ pixels.}
\label{fig08}
\end{figure}

Second, we choose another situation where the PSF is larger than a pixel of the detector.  A Gaussian function with
$\sigma=10$ pixels  is set as the PSF of the imaging system (see the top plot in Fig.\ \ref{fig07}).
100 simulated observed images are generated; one of them is shown in Fig.\ \ref{fig07} (top plot). 
The relevant CC matrix is shown in Fig.\ \ref{fig08}. Three coefficient curves
related to specific pixels 120, 160 and 330 are extracted from that matrix, as shown in Fig.\ \ref{fig07} (bottom plot).

In the second situation,  there are non-uniform extended structures in the CC matrix (see Fig.\ \ref{fig08}), 
which are induced by the imaging device. 
Fig.\ \ref{fig07} and \ref{fig08} show that the shape of  CC curves depends on the structures of 
the object, on the PSF and on the noise level.

\subsection{Statistical Definition of Image Resolution} \label{sec3B}
\begin{figure}[!h]
\centering
\includegraphics[width=\textwidth]{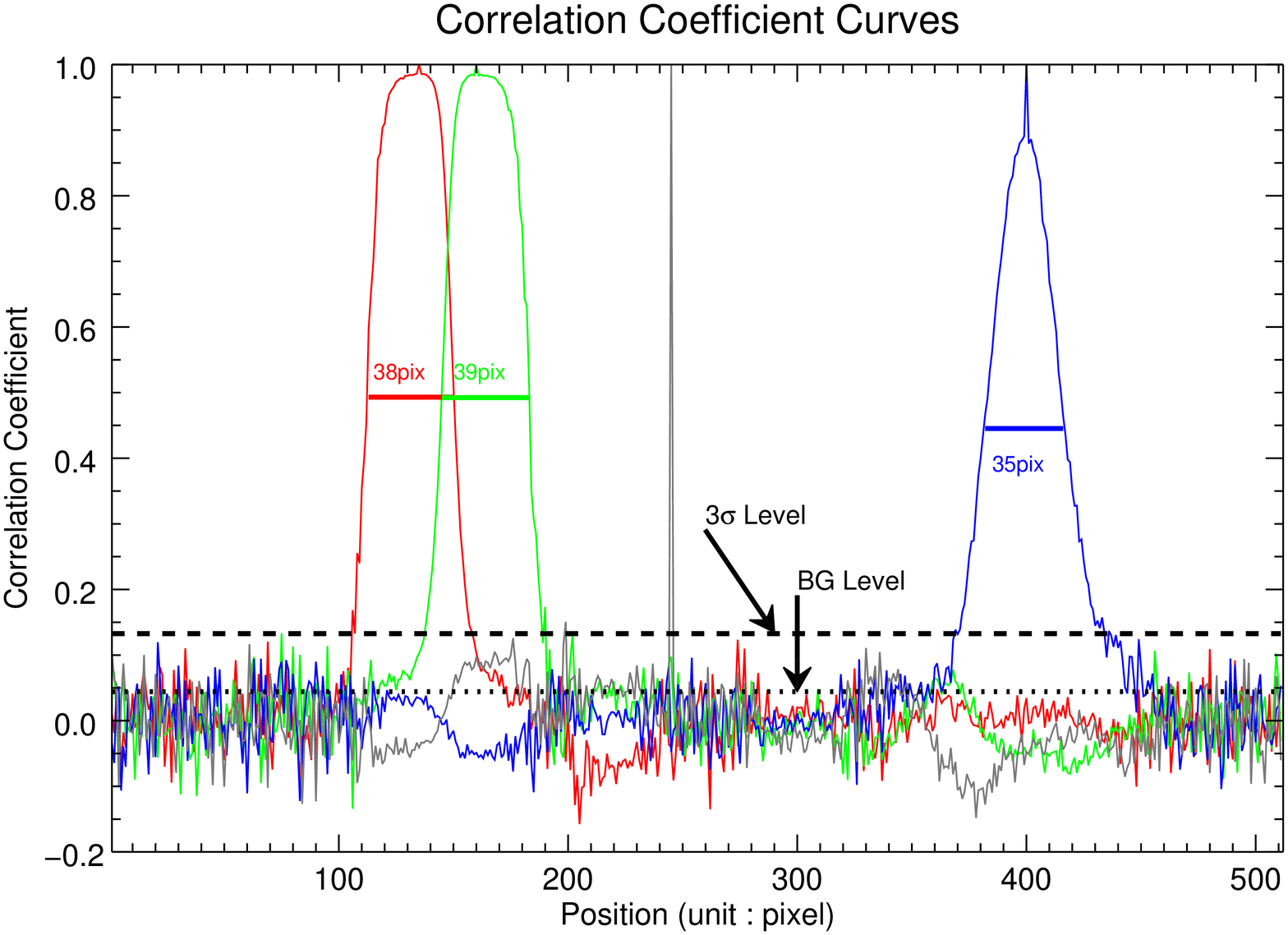}
\caption{An example of image resolution based on the correlation of pixels. 
Four CC curves related to pixels 120 (red), 160 (green), 245 (grey) and 330 (blue) respectively
are extracted from the CC matrix. Background  (BG) level (dotted line) and $3~\sigma$ level (dashed line)
are shown in the figure too.  The detailed parameters of this example are described in 
the third paragraph of  Section \ref{sec3B}. }
\label{fig10}
\end{figure}

A statistical definition of angular resolution is quite straightforward. A correlation coefficient (CC) matrix or tensor
 could be
derived from a series of observed or deconvolved images. Next, one needs to find the background level of CCs. 
If there is signal in the images, the CLEAN (\cite{hogbom1974aperture}) algorithm could be 
used to remove this signal from the images. Then, a CC matrix can be calculated from the
residual images where only incoherent noise exists.  
The standard deviation of the CC matrix without its diagonal elements can be calculated.   
By setting a confidence
level (e.g., $3~\sigma$ level ) above which the correlations  between pairs of pixels are believed to be real, 
we are able to  study the resolution properties of the images(see Fig.\ \ref{fig10}).

A quantitive definition of resolution can be derived from CC curves or matrices. If the structure 
of the CC curves or matrices is simple, just like those shown in Fig.\ \ref{fig10}, the FWHMs of the curves
can define the  resolution at the  corresponding pixels. Sometimes, the structure of CC curves is quite 
complex, as indicated by the examples in Section \ref{sec4}. In this situation, it is possible to define
the resolution as the FWHM of the main beams of CC curves.

Let's use an example to demonstrate the idea of an image resolution based on the pixel correlations , with
an image system that has a Gaussian PSF with $\sigma = 10\,\mathrm{px}$. 
The object or true image consists of four point sources and one extended source.
The point sources are at positions of $120\,\mathrm{px}$, $130\,\mathrm{px}$, $220\,\mathrm{px}$ and $300\,\mathrm{px}$, with brightness of $8.0\times 10^4$\,counts/s, 
$8.0\times 10^4$\,counts/s, $2.0\times 10^4$\,counts/s and $4.0\times 10^4$\,counts/s respectively. The extended source ranging from $340\,\mathrm{px}$ to $440\,\mathrm{px}$ has a
mean brightness of $5.0\times 10^3$\,counts/s. The additive noise is also Gaussian, with $\sigma=2.0$\,counts/s.
Four CC curves are extracted from the CC matrix and are shown in Fig.\ \ref{fig10}.

As indicated in Fig.\ \ref{fig10}, the resolution in an image  is variable.  The CC curves related to pixel 120 and 160
has FWHM of 38 pixel and 39 pixel respectively. The CC curve related to pixel 330 has FWHM of 35 pixel. At pixel 245, the CC curve is a $\delta$ function since there is only noncoherent noise, which shows that resolution is 
 meaningless there.



\section{Image Resolution in Deconvolution} \label{sec4}
Nowadays, there are many deconvolution algorithms, such as CLEAN (\cite{hogbom1974aperture}), Maximum Likelihood (\cite{akaike1998information}),
Wiener filter (\cite{wiener1949extrapolation}), (Kalman filter\cite{kalman1960new}), 
Maximum Entropy (\cite{jaynes1957information}), Maximum A Posteriori (\cite{degroot2005optimal}) , etc. 
All these methods intend to
provide a good estimation of the original objects. In some situation, for example with data of good quality, 
deconvolution can produce images with super-resolution. The super-resolution, as mentioned before, can 
not be described by the classical definition of resolution. 

With the new definition of image resolution presented in Section \ref{sec3B}, however, the super-resolution can be  
quantatively evaluated.  We use the Wiener deconvolution for our demonstration. 
Other deconvolution algorithms are also applicable.

\subsection{Wiener Deconvolution}
Wiener deconvolution, which is based on the Wiener filter, is a good algorithm for removing
the effects of a point spread function.  The main steps of the algorithm are 
reviewed below.

For an imaging system described in (\ref{imageconv}), 
the goal of the Wiener deconvolution is to find a $\mathbf{g}$ so that it is possible to estimate $\mathbf{o}$ 
by following operation:
\begin{equation}
\hat{\mathbf{o}} = \mathbf{g} \otimes \mathbf{d}
\end{equation}
where $\hat{\mathbf{o}}$, which minimizes the mean square error,  is an estimate of $\mathbf{o}$. 
The solution $\mathbf{g}$ can be more easily expressed in the frequency domain : 
\begin{equation}
\mathbf{G}(f) = \frac{\mathbf{P}(f)^* \mathbf{O}(f)}{|\mathbf{P}(f)|^2 \mathbf{O}(f) + \mathbf{N}(f)}
\end{equation}
where $\mathbf{G}$ and $\mathbf{P}$ are the Fourier transforms of $\mathbf{g}$ and 
$\mathbf{p}$ respectively, $\mathbf{O}$ and 
$\mathbf{N}$ are the mean power spectral density of $\mathbf{o}$ and $\mathbf{n}$, the superscript $*$ denotes
complex conjugation, $f$ denotes frequency.

\subsubsection{Example I: Low Signal-to-Noise Ratio}  \label{LowSNR}
\begin{figure}[!h]
\centering
\includegraphics[width=\textwidth]{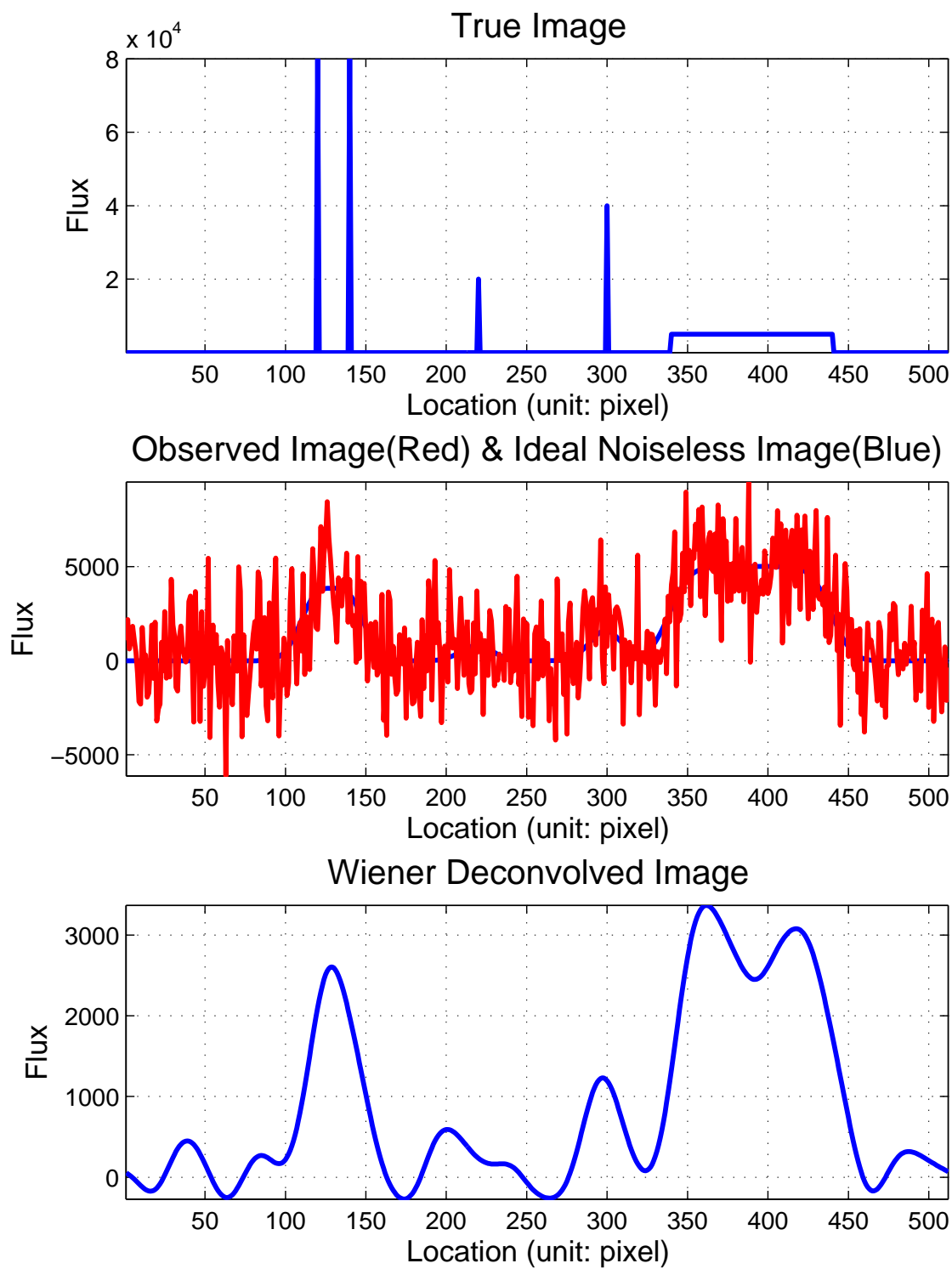}
\caption{The true image (top plot), the observed image as well as the ideal noiseless image (middle plot) and
 the Wiener deconvolved image (bottom plot) in Example I where the SNR of the observed image is about 1.3.}
\label{fig11}
\end{figure}

In this example, we use the same object (true image) as that of Section \ref{sec3b}. a Gaussian point spread function with $\sigma = 10\,\mathrm{px}$ is used. 
The object or true image is shown  in Fig.\ \ref{fig11} (top plot).
100 observed images were generated. Each image is  the convolution of a 
Poisson sample of the original object with the PSF, whose FWHM is about $23\,\mathrm{px}$, 
plus additive  Gaussian noise with 
$\sigma=2000$\,counts/s.  The corresponding SNR is  about 1.3. One of the observed images is shown in Fig.\ \ref{fig11} (middle plot). A Wiener deconvolution was applied to each sample of observed image. 
One deconvolved image is shown in Fig.\ \ref{fig11}  (bottom).  
 
\begin{figure*}[!h]
\centering
\includegraphics[width=\textwidth]{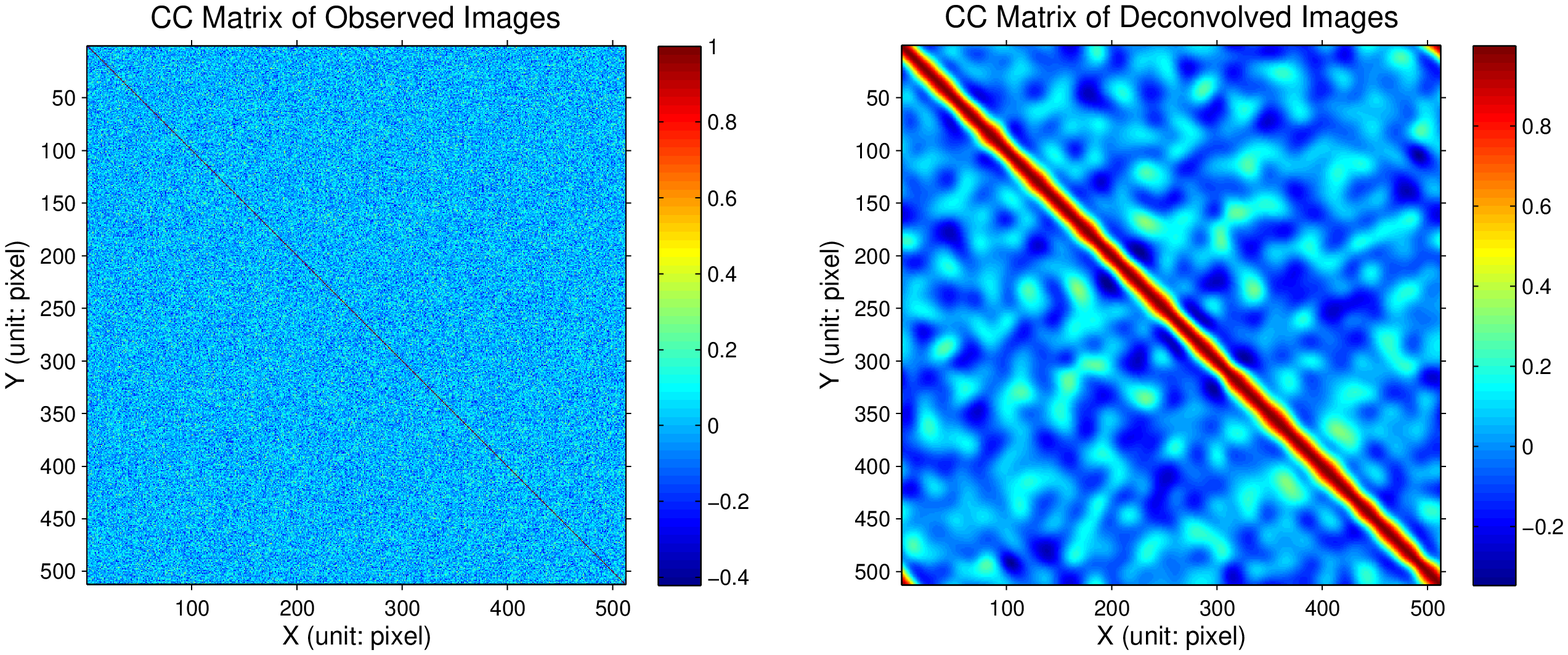}
\caption{CC matrices of observed images (left plot) and deconvolved images (right plot).  The SNRs of 
the observed images are about 1.3}
\label{fig12}
\end{figure*}

\begin{figure}[!h]
\centering
\includegraphics[width=\textwidth]{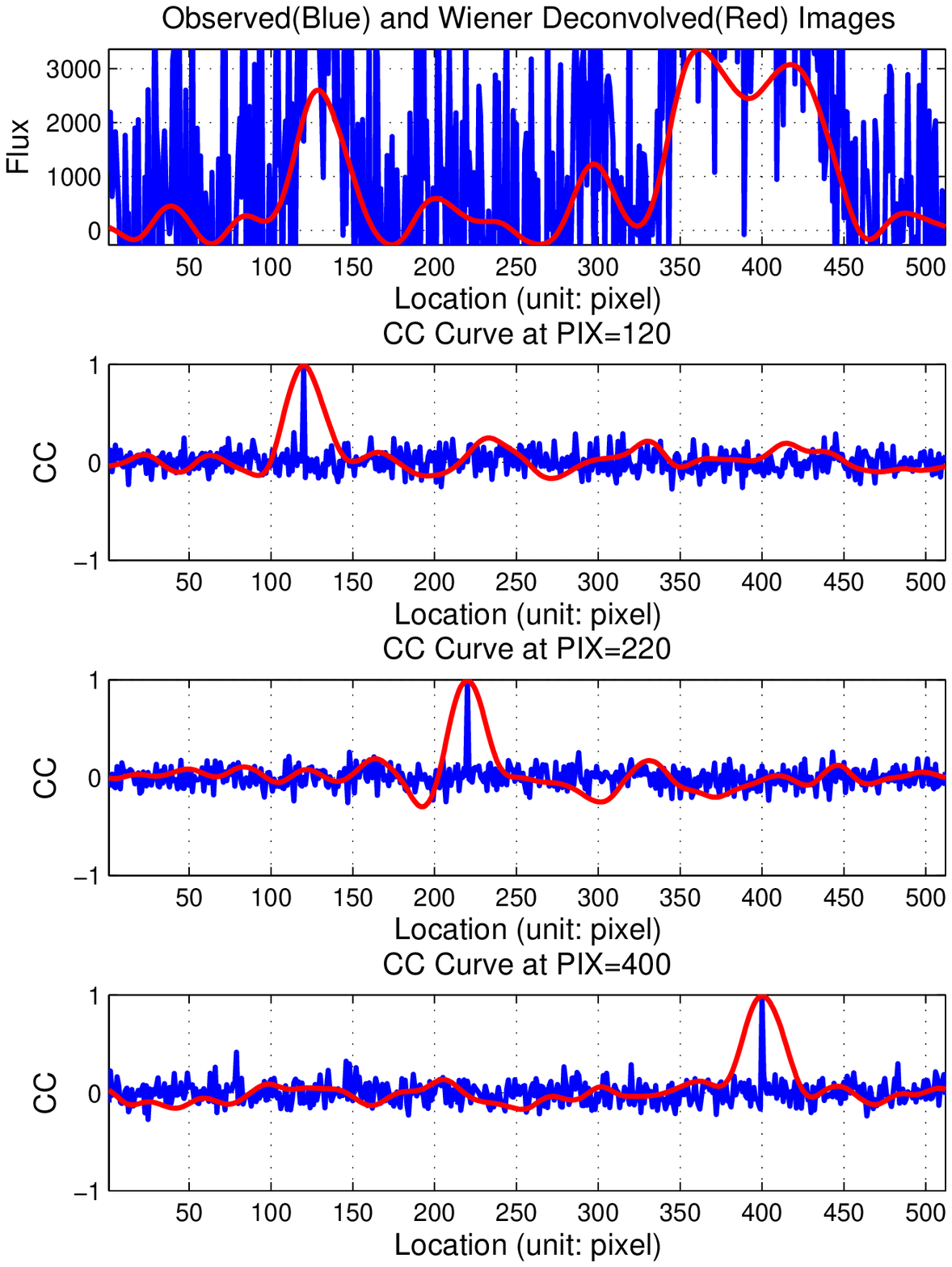}
\caption{Brightness distributions of an observed (blue) image and of its Wiener-deconvolved (red) image (top plot). 
The other plots show three pairs of CC curves , both for the observed (blue) and  Wiener-deconvolved (red) image, for pixels $120\,\mathrm{px}$,
 $220\,\mathrm{px}$ and $400\,\mathrm{px}$ respectively.}
\label{fig13}
\end{figure}

We produced 100 samples of observed images and Wiener deconvolved images, their CC matrices 
(see Fig.\ \ref{fig12}) were calculated, and CC curves at different positions (see Fig.\ \ref{fig13})
 were extracted. In this situation, the noise is strong and has no correlation. In the CC matrix, there are 
only diagonal elements whose values equal 1.0 and with the non-diagonal elements very close  to 0.0 
(see Fig.\ \ref{fig12}, left plot).  Therefore, there is no meaningful image resolution here.

The Wiener deconvolution is capable of suppressing noise and of detecting weak sources, as shown  in Fig.\ \ref{fig11}
and Fig.\ \ref{fig13}.  After deconvolution, an estimation of the object is produced, and  
a structure appears in the  CC matrix (see Fig.\ \ref{fig12}, right plot, and Fig.\ \ref{fig13}). The FWHMs of the main beams in the CC
curves are listed in Table \ref{fwhm}.  The image resolution here displays small variation, and 
is comparable to the FWHM of the Gaussian PSF, which is about $23\,\mathrm{px}$. 

\subsubsection{Example II : High Signal-to-Noise Ratio} \label{HighSNR}
In this example, again, we use the same object (true image) as that of Section \ref{sec3b}. 
The Gaussian noise was reduced to $\sigma=2.0$\,counts/s. The corresponding SNR is about 930. 

\begin{figure}[!h]
\centering
\includegraphics[width=\textwidth]{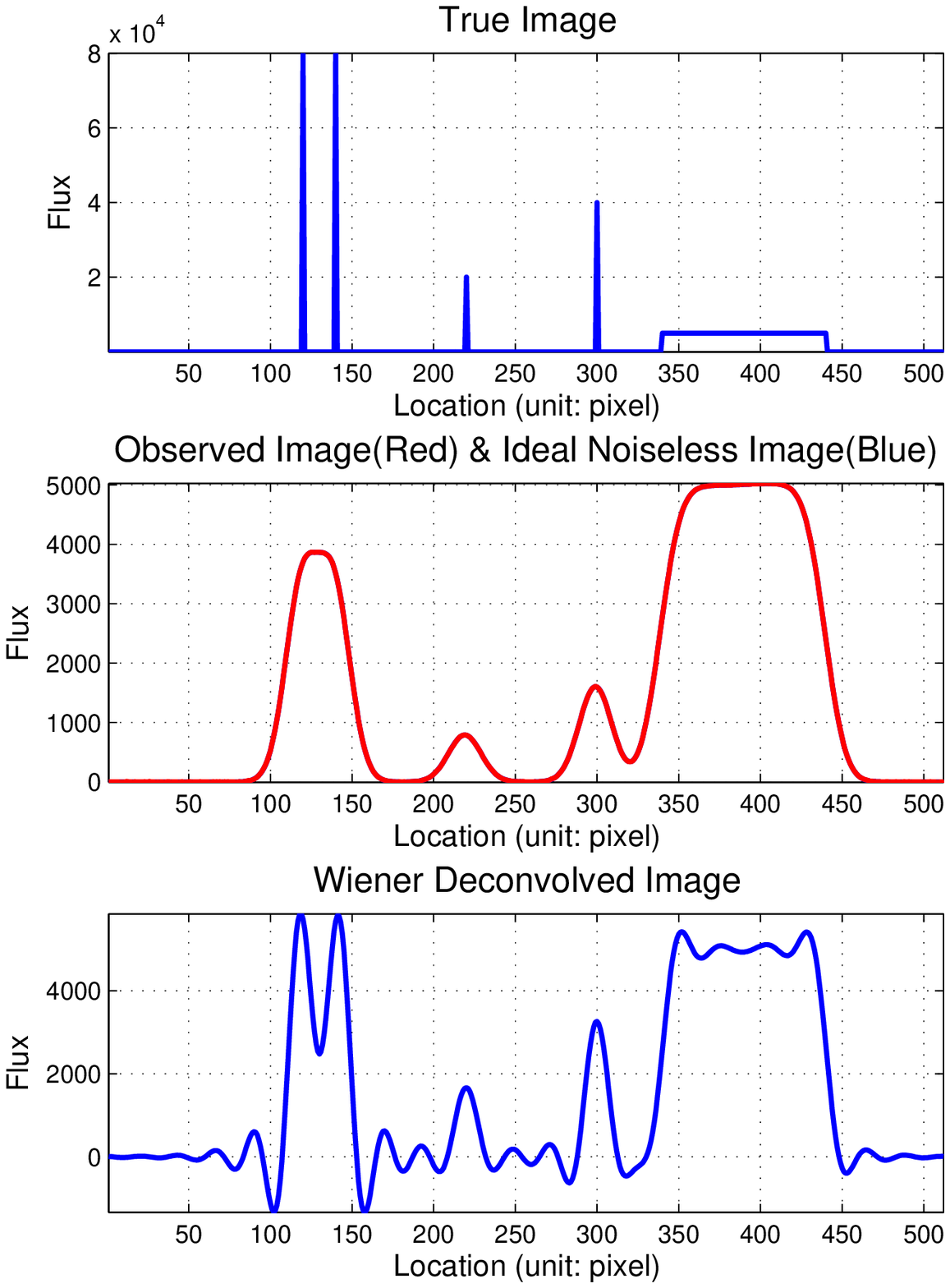}
\caption{The true image (top plot), the observed image as well as ideal noiseless image (middle plot) and
 Wiener deconvolved image (bottom plot) in Example II where the SNR of the observed image is about 930.}
\label{fig14}
\end{figure}

\begin{figure*}[!h]
\centering{
\includegraphics[width=\textwidth]{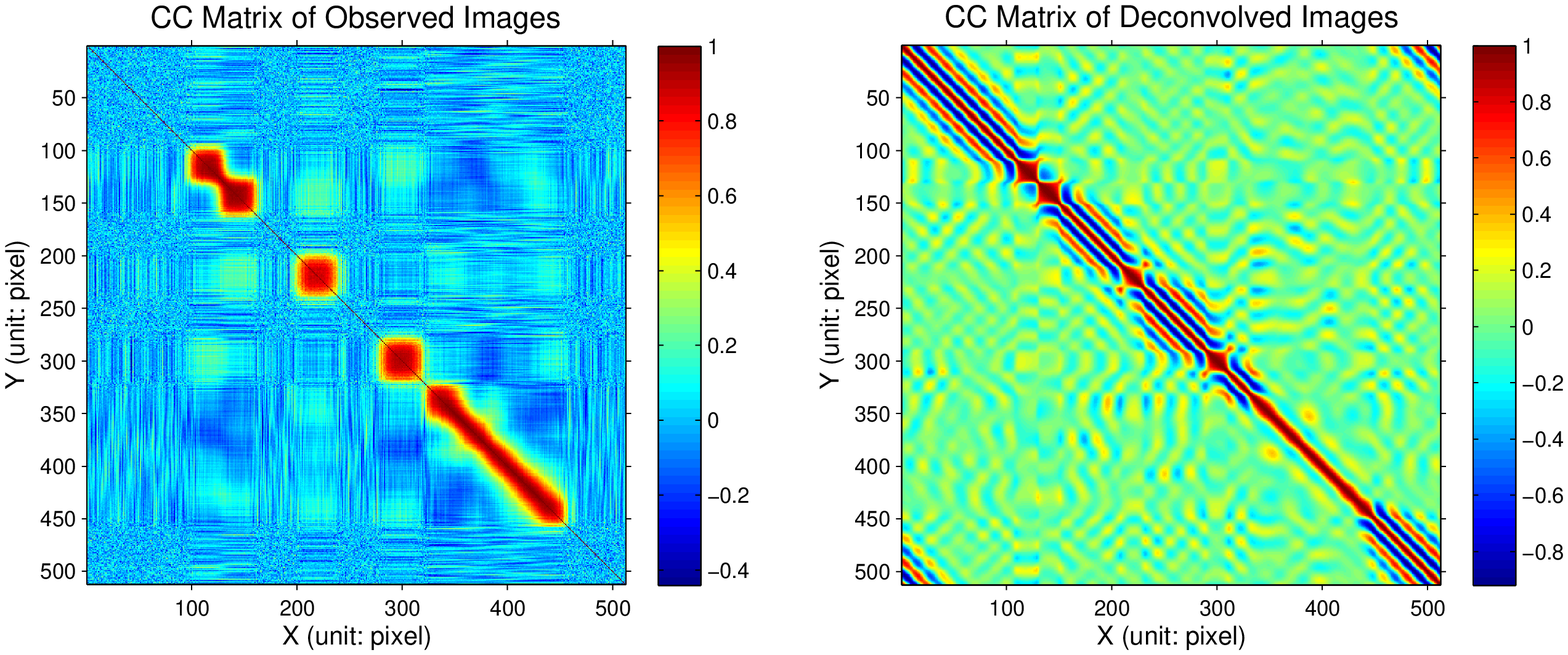}
\caption{The CC matrices of observed images (left plot) and deconvolved images (right plot).  
The SNRs of the observed images are about 930.}
 \label{fig15} }
\end{figure*}

\begin{figure}[!ht]
\centering
\includegraphics[width=\linewidth]{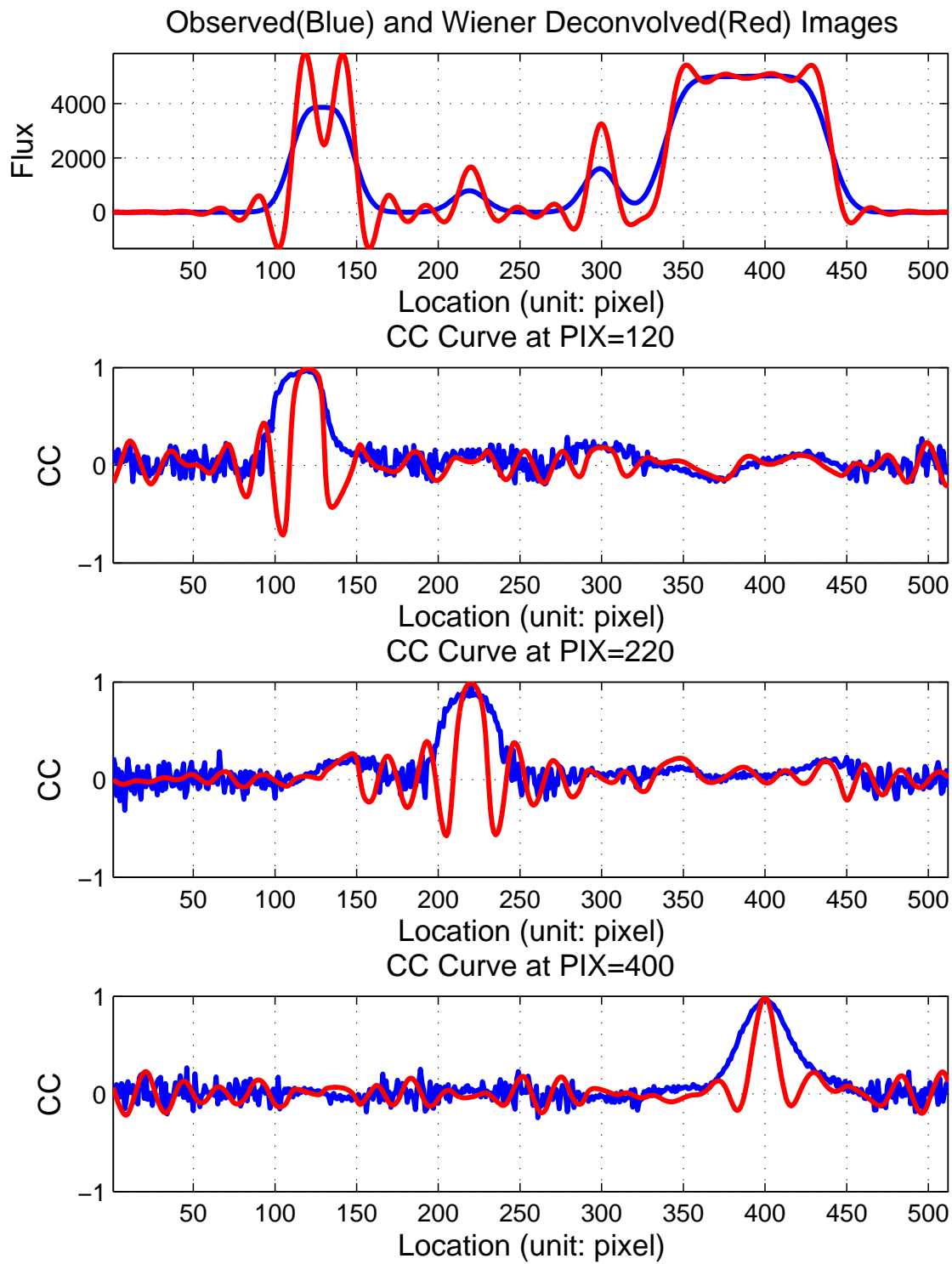}
\caption{The brightness distributions of an observed(blue) and a Wiener deconvolved(red) images (the 1st plot). 
Three pairs of CC curves (the 2nd, 3rd and 4th plots) related to observed (blue) and Wiener deconvolved (red) images at positions $120\,\mathrm{px}$,
 $220\,\mathrm{px}$ and $400\,\mathrm{px}$ respectively. }
\label{fig16}
\end{figure}

Fig.\ \ref{fig14} shows the object or true image (top plot), one of the observed image
(middle plot) and its Wiene- deconvolved image (bottom plot).  Fig.\ \ref{fig15} shows the CC matrices
of 100 observed images (left plot) and of their Wiener deconvolved images (right plot).  Three pairs of CC curves at positions $120\,\mathrm{px}$,
 $220\,\mathrm{px}$ and $400\,\mathrm{px}$ respectively are shown in Fig.\ \ref{fig16}.

We see in Fig.\ \ref{fig15} and \ref{fig16} and Table \ref{fwhm} that  the image resolution of observed 
images ranges from $32\,\mathrm{px}$ to $35\,\mathrm{px}$ is larger than the FWHM of the PSF, which is 
about $23\,\mathrm{px}$. How the SNR affects the image resolution will be discussed in Section \ref{sec5}. The table  also shows that the resolution after Wiener deconvolution  ranges from $14.0\,\mathrm{px}$ to $19.0\,\mathrm{px}$, 
which is better than the resolution of the observed images. The resolution is also less than
the FWHM of the PSF,  which properly indicates that  Wiener deconvolution achieves super-resolution.
Looking into the bottom plot in Fig.\ \ref{fig14}, we can see that two point sources at $120\,\mathrm{px}$ and 
$140\,\mathrm{px}$ are clearly resolved,  although the distance between them is less than the FWHM of the PSF.
 
\begin{table}[!t]
\renewcommand{\arraystretch}{1.3}
\caption{The FWHMs (in pixels) of the Main Beams of CC Curves in the Wiener Deconvolution}
\label{fwhm}
\centering
\begin{tabular}{l|cc|cc}
\hline\hline
Position & \multicolumn{2}{|c|}{SNR=1.3} & \multicolumn{2}{c}{SNR=930} \\
& (Observed) & (Deconvolved) & (Observed) & (Deconvolved) \\
\hline
120\,px & N/A  & 22.0 &  35.0 & 19.0 \\
220\,px & N/A  & 25.0 &  32.0 & 16.0 \\
400\,px & N/A  & 24.0 &  34.0 & 14.0\\
\hline
\end{tabular}
\end{table}

\section{Discussion} \label{sec5}
To the first order approximation, an imaging system can be described by  (\ref{imageconv}).  
In the new definition of resolution given in this paper  assumes that the objects are variable. 
Their variations, after convolution with a PSF,  induces correlation in an observed
image. 

In a real imaging system, quantum mechanics controls the behavior of photons. 
A photon arrives at a pixel of a detector with a  probability  which is dominated by the  normalized profile of a PSF.
So, there will be no correlation in a series of real observed images. An exception is if objects are variable.
Simulation shows that if a point source has a fluctuation which is larger than its intrinsic fluctuation ($1/\sqrt{N}$), 
then the cross-correlation coefficient curve has the  shape of the PSF.

Even though the new definition in this paper doesn't  describe the quantum nature of a real imaging system, 
it is still a powerful analytical tool that can be used for evaluating an imaging system in some special objects under 
some special noise levels, and also for assessing the performance of a deconvolution algorithm.

As shown in the examples of Section \ref{sec4}, for observed images, the higher the SNR the worse the
image resolution. This may be contrary to our intuition, but the reason is simple. For bright sources, their
influence on nearby object is strong, through convolution with  a PSF, and the additive noise is weak. Therefore,
the main beams of their CC curves will be broader.  For high-quality observed images with high SNR, 
deconvolution can help improve the resolution, or 
even reach super-resolution results, as shown in Example~II.

\section{Conclusions} \label{sec6}
In this paper, a new definition of image resolution based on the correlation between pixels was introduced. 
For a series of observed images, the correlation is mainly due to the point spread functions of the imaging
system, but is also influenced by the structures of the object and by any additive noise.  

The resolution may vary across an image. The new definition is also suitable for the interpretation of deconvolution results. The Wiener deconvolution was use in our demonstration. 
It was found that weak structures can be extracted from low signal-to-noise ratio data, but with low resolution; on the other hand super resolution image can be obtained from high signal-to-noise ratio data after Wiener deconvolution. The new definition can also be used to compare various deconvolution algorithms on the effect of their processing: resolution, sensitivity and sidelobe level etc.  

\section*{Acknowledgments}
This work was supported by the National Natural Science Foundation of China (NSFC) under Grant
No.\ 11173038 and No.\ 11373025, and also by the Tsinghua University Initiative Scientific Research Program under Grant No.\
20111081102. The author would like to thank Dr. Eric O. Lebigot for his kind help for the English improvement of this paper.


\begin{thebibliography}{99}
\bibitem[Akaike 1998]{akaike1998information}
Akaike H.,  1998, in Selected Papers of Hirotugu Akaike, Springer, p. ~199

\bibitem[Banterle et al. 2013]{banterle2013fourier}
Banterle N., Bui K. H., Lemke E. A., and Beck M., 2013, Journal of
  structural biology, 183, 363

\bibitem[Blanter \& B{\''u}ttiker 2000]{blanter2000shot}
Blanter Y. M.,  B{\"u}ttiker M., 2000, Physics reports, 336, 1
    
\bibitem[Born \& Wolf 1999]{born1999principles}
Born M.,  Wolf E., 1999, Principles of optics: electromagnetic theory of
  propagation, interference and diffraction of light, CUP Archive

\bibitem[Conchello 1998]{conchello1998superresolution}
Conchello J. A., JOSA A, 1998, 15, 2609

\bibitem[DeGroot 2005]{degroot2005optimal}
DeGroot M. H., 2005, Optimal statistical decisions, John Wiley \& Sons, 82
  
\bibitem[Fried 1966]{fried1966optical}
Fried D. L., 1966, JOSA, 56, 1372

 \bibitem[H{\''o}gbom 1974]{hogbom1974aperture}
H{\"o}gbom J., 1974,  Astron. Astrophys. Suppl, 15, 417
         
\bibitem[1991]{irani1991improving}
Irani M.,  Peleg S., 1991, CVGIP: Graphical models and image processing, 53, 231

\bibitem[Jaynes 1957]{jaynes1957information}
Jaynes E. T., 1957, Physical review, 106, 620
  
\bibitem[Janesick 2001]{janesick2001scientific}
Janesick J. R., 2001, Scientific charge-coupled devices, SPIE press Bellingham, Washington, 117
 
\bibitem[Kalman 1960]{kalman1960new}
Kalman R. E., 1960, Journal of Fluids Engineering, 82, 35
     
\bibitem[Labeyrie 1970]{labeyrie1970attainment}
Labeyrie A., 1970, A \& A, 6, 85

\bibitem[MacDonald 2006]{macdonald2006digital}
MacDonald L. W., 2006, Digital heritage: applying digital imaging to cultural
  heritage, Routledge
  
\bibitem[Nieuwenhuizen et al.  2013]{nieuwenhuizen2013measuring}
Nieuwenhuizen R. P., Lidke K. A., Bates M., Puig D. L., Gr{\"u}nwald D.,
  Stallinga S.,  Rieger B., 2013, Nature methods, 10, 557

\bibitem[Pearson 1985]{pearson1895note}
Pearson K., 1895, Proceedings of the Royal Society of London, 58, 240
      
\bibitem[Saxton \& Baumeister 1982]{saxton1982correlation}
Saxton W.,  Baumeister W., 1982, Journal of Microscopy, 127, 127

\bibitem[Tuzlukov 2010]{tuzlukov2010signal}
Tuzlukov V., 2010, Signal processing noise, CRC Press, 8
  
\bibitem[Ubertini et al. 2003]{ubertini2003ibis}
Ubertini P., Lebrun F. et al., 2003, A \& A, 411, 131

\bibitem[Wiener 1949]{wiener1949extrapolation}
Wiener N., 1949, Extrapolation, interpolation, and smoothing of stationary time
  series. MIT press Cambridge, MA, 2

\end{thebibliography}
\end{document}